\documentclass[twocolumn,times,astrosymb]{aastex63111}

\usepackage{amsmath}
\usepackage{upgreek}
\usepackage{tikz}
\usepackage{comment}

\newcommand{\burst}{burst}

\newcommand{\nusedant}{19}
\newcommand{\obsdate}{2024 June $13^{\rm th}$}
\newcommand{\utc}{05:13:09.035\,UTC}
\newcommand{\mjd}{60474.21746568846}
\newcommand{\tres}{13.8\,ms} 
\newcommand{\totalduration}{$30$\,ns}
\newcommand{\duration}{$10$\,ns}
\newcommand{\detsnr}{16.3}
\newcommand{\detbeam}{17} 

\newcommand{\bufferflux}{112\,kJy} 
\newcommand{\flux}{300\,kJy} 
\newcommand{\fluxmax}{3\,MJy} 
\newcommand{\dm}{$2.26 \cdot 10^{-5}$\,\pccc}
\newcommand{\tecu}{69.7\,TECU}

\newcommand{\rtfmin}{760.5\,MHz}
\newcommand{\rtfmax}{1000.5\,MHz}

\newcommand{\bfmin}{695.5\,MHz}
\newcommand{\bfmax}{1031.5\,MHz}
\newcommand{\ffra}{1h34m35.6s $\pm$ 8.1s}
\newcommand{\ffdec}{-10$^{\circ}$24'27.6" $\pm$ 3.7"}
\newcommand{\dist}{4500\,km}
\newcommand{\disterr}{80\,km}

\newcommand{\sat}{Relay 2} 

\newcommand{\pccc}{\ensuremath{{\rm pc\,cm^{-3}}}}

\shorttitle{Nanosecond satellite ESD}
\shortauthors{James et al.}
\graphicspath{{./}{figs/}}

\begin{document}

\title{A nanosecond-duration radio pulse originating from the defunct Relay 2 satellite}

\author[0000-0002-6437-6176]{C.~W.~James}
\altaffiliation{E-mail: clancy.james@curtin.edu.au}\affiliation{International Centre for Radio Astronomy Research (ICRAR), Curtin University, Bentley, WA 6102, Australia}


\author[0000-0001-9434-3837]{A.~T.~Deller}
\affiliation{Centre for Astrophysics and Supercomputing, Swinburne University of Technology, John St., Hawthorn, VIC 3122, Australia}

\author[0009-0004-1205-8805]{T.~Dial}
\affiliation{Centre for Astrophysics and Supercomputing, Swinburne University of Technology, John St., Hawthorn, VIC 3122, Australia}

\author[0000-0002-5067-8894]{M.~Glowacki}\affiliation{International Centre for Radio Astronomy Research (ICRAR), Curtin University, Bentley, WA 6102, Australia}
\affiliation{Institute for Astronomy, University of Edinburgh, Royal Observatory, Edinburgh, EH9 3HJ, United Kingdom}
\affiliation{Inter-University Institute for Data Intensive Astronomy, Department of Astronomy, University of Cape Town, Cape Town, South Africa}

\author[0000-0002-8195-7562]{S.~J.~Tingay}
\affiliation{International Centre for Radio Astronomy Research (ICRAR), Curtin University, Bentley, WA 6102, Australia}


\author[0000-0003-2149-0363]{K.~W.~Bannister}
\affiliation{Australia Telescope National Facility, CSIRO, Space and Astronomy, PO Box 76, Epping, NSW 1710, Australia}
\affiliation{Sydney Institute for Astronomy, School of Physics, University of Sydney, NSW 2006, Australia}

\author[0000-0002-2864-4110]{A.~Bera}
\affiliation{International Centre for Radio Astronomy Research (ICRAR), Curtin University, Bentley, WA 6102, Australia}

\author[0000-0002-8383-5059]{N.~D.~R.~Bhat}
\affiliation{International Centre for Radio Astronomy Research (ICRAR), Curtin University, Bentley, WA 6102, Australia}

\author[0000-0002-3532-9928]{R.~D.~Ekers}
\affiliation{Australia Telescope National Facility, CSIRO, Space and Astronomy, PO Box 76, Epping, NSW 1710, Australia}
\affiliation{International Centre for Radio Astronomy Research (ICRAR), Curtin University, Bentley, WA 6102, Australia}

\author[0000-0002-4382-0100]{V.~Gupta}
\affiliation{Australia Telescope National Facility, CSIRO, Space and Astronomy, PO Box 76, Epping, NSW 1710, Australia}

\author[0000-0002-8987-1544]{A.~Jaini}
\affiliation{Centre for Astrophysics and Supercomputing, Swinburne University of Technology, Hawthorn, VIC 3122, Australia}

\author[0000-0001-9224-5483]{J.~Morgan}
\affiliation{CSIRO Space and Astronomy, P.O. Box 1130, Bentley, WA 6102, Australia}

\author[0000-0003-4193-6158]{J.~N.~Jahns-Schindler}
\affiliation{Centre for Astrophysics and Supercomputing, Swinburne University of Technology, Hawthorn, VIC 3122, Australia}
\affiliation{ARC Centre of Excellence for Gravitational Wave Discovery (OzGrav), Hawthorn, VIC 3122, Australia}

\author[0000-0002-7285-6348]{R.~M.~Shannon}
\affiliation{Centre for Astrophysics and Supercomputing, Swinburne University of Technology, Hawthorn, VIC 3122, Australia}
\affiliation{ARC Centre of Excellence for Gravitational Wave Discovery (OzGrav), Hawthorn, VIC 3122, Australia}

\author[0009-0004-6311-2859]{M.~Sukhov}
\affiliation{School of Science, Royal Melbourne Institute of Technology, Melbourne, VIC, 3000}

\author[0000-0002-7551-2073]{J.~Tuthill}
\affiliation{Australia Telescope National Facility, CSIRO, Space and Astronomy, PO Box 76, Epping, NSW 1710, Australia}

\author[0000-0002-2066-9823]{Z.~Wang}
\affiliation{International Centre for Radio Astronomy Research (ICRAR), Curtin University, Bentley, WA 6102, Australia}

\begin{abstract}

We report the detection of a burst of emission over a \bfmin--\bfmax\ bandwidth by the Australian Square Kilometre Array Pathfinder, ASKAP. The burst was localised through analysis of near-field time delays to the long-decommissioned \sat\ satellite, and exhibited a dispersion measure of \dm\ --- \tecu, consistent with expectations for a single pass through the ionosphere. After coherent dedispersion, the \burst\ was determined to be less than \totalduration\ in width, with an average flux density of at least \flux. We consider an electrostatic discharge (ESD) or plasma discharge following a micrometeoroid impact to be plausible explanations for the \burst. ESDs have previously been observed with the Arecibo radio telescope, but on 1000 times longer timescales. Our observation opens new possibilities for the remote sensing of ESD, which poses a serious threat to spacecraft, and reveals a new source of false events for observations of astrophysical transients.
\end{abstract}


\section{Introduction}
\label{sec:intro}

The charging of spacecraft in orbit due to interactions with the space environment has been a well-known phenomena since the early days of the space program \citep{SpacecraftCharging1972}. Accumulation of electrons and ions can lead to large voltage differentials between spacecraft surfaces, and between the spacecraft and space plasmas. Sudden ESD in the form of arc currents can cause damage to spacecraft systems, and although decades of experience has led to design recommendations to avoid this \citep{MitigatingChargingEffects}, spacecraft charging remains a hazard. A recent likely example is the temporary loss of the Galaxy 15 satellite in April 2010 \citep{2010SpWea...8.6008A}, while numerous, minor discharges have been proposed for the gradual degradation of solar arrays supplying power to GPS satellites \citep{FergusenGPS2016}. Despite this, most spacecraft do not contain mechanisms for monitoring the build-up of charge, or detecting sudden discharges, and the primary method of detection remains the study of the damage caused by ESD \citep[e.g.\ ][]{1995STIN...9611547L}.

The remote sensing of ESD by low-frequency ground-based radio monitors has been proposed as a method to study this phenomena \citep{2014JSpRo..51.1907F}. Strong $\sim$MHz radio bursts with durations of microseconds have been observed in laboratory experiments aimed at reproducing ESD from the charging of solar arrays \citep{1985jpl..rept.....L}, and recent observations with the Arecibo telescope have detected 300--350\,MHz emission typically lasting 40--50\,$\upmu$s from GPS satellites \citep{AreciboGPS2017}. However, there has been no systematic campaign  capable of observing large numbers of spacecraft.

It has also been proposed that the energetic plasmas produced by high-velocity micrometeoroid impacts with spacecraft may produce radio-frequency emission \citep{2013ITPS...41.3545G}. However, like ESD, literature expectations for this emission suggests durations of at least several $\upmu$s, and emission peaked at low MHz or kHz frequencies \citep{LI2023104619}.

Here we report the serendipitous detection of a \totalduration\ duration, \flux\ impulse from the decommissioned \sat\ satellite by the Australian Square Kilometre Array Pathfinder \citep[ASKAP; ][]{Hotan2021ASKAP}, which challenges expectations for both ESD and micrometeoroid origins.

We discuss the signal in \S\ref{sec:detection}, the origin in the \sat\ satellite in \S\ref{sec:origin}, and the implications for both remote sensing of ESD, and false event rates for astroparticle experiments studying nanosecond radio pulses, in
\S\ref{sec:implications}.

\section{Detection}
\label{sec:detection}

The \burst\ was detected on \obsdate\ at \utc\, (i.e., MJD \mjd) by ASKAP using the Commensal Realtime ASKAP Fast Transients Survey (CRAFT) coherent detection system, CRACO \citep{Andy2024CRACO}. The CRACO system uses aperture synthesis techniques to form  total intensity (Stokes I) images in near-real-time, using data from \rtfmin\ to \rtfmax, in 1\,MHz channels, at \tres\ resolution. ASKAP uses phased-array feed technology to form 36 contiguous images on the sky, each of which are processed separately in parallel by the real-time CRACO system.
The \burst\ was detected in beam \detbeam\ of ASKAP's 36 beams with a signal-to-noise ratio of \detsnr\ at a time resolution of \tres, with no discernible dispersion in frequency.

Most FRB detection pipelines --- including earlier FRB searches performed by ASKAP \citep{shannon24craftics} ---  routinely reject signals with low dispersion measure due to the prevalence of transient radio-frequency interference (RFI). However, the coherent imaging used by the ASKAP/CRACO system has much stronger terrestrial RFI rejection;  terrestrial RFI has a very significantly curved wavefront across the array and will not form a point-source in the image plane, unlike a far-field astrophysical source, facilitating the identification and removal of candidates generated by it. We were additionally motivated to search for undispersed bursts because Galactic objects such as rotating radio transients and long-period transients are detectable by the CRACO system \citep{Andy2024CRACO}, and may appear undispersed at our time-resolution of \tres\ if they have a DM of $\lesssim 10$\,\pccc.

Our detection triggered a download of 12\,s of voltage data (which represents the electric field as sampled by each ASKAP antenna) surrounding the event, for a band from \bfmin\ to \bfmax\ in complex-valued 1 MHz subbands. The raw voltage data saved in this fashion have been quantised with 1-bit precision prior to storage in the buffer, to provide a buffer of useful time length given the available memory; the real-time CRACO detection system makes use of the data prior to this re-quantisation stage, which has much higher bit depth. The downloaded voltage data were then run through ASKAP's offline processing pipeline \citep[CELEBI;][]{Scott2023_CELEBI} to localise the burst.

Iterative offline processing revealed that only antennas within the inner 1\,km of ASKAP produced a coherent image under the usual assumption of a far-field source whose wavefronts are planar as they pass the telescope array. Using this sub-set of the array, we recovered a point-like source with apparent position RA=\ffra, DEC=\ffdec. As longer baseline data was included, the source S/N progressively reduced.  This is indicative of an origin within the near-field limit of ASKAP, which is $2 D^2 \lambda^{-1} \sim 200,000$\,km for the full ASKAP baseline of $D=6$\,km, but only 5700\,km for $D=1$\,km. The burst was unresolved at the standard minimum CELEBI processing timescale of $1\,\upmu$s. These characteristics initiated a dedicated analysis chain for this event.

\begin{figure}
\centering
\includegraphics[width=\columnwidth,trim={0cm 0cm 0cm 0cm},clip]{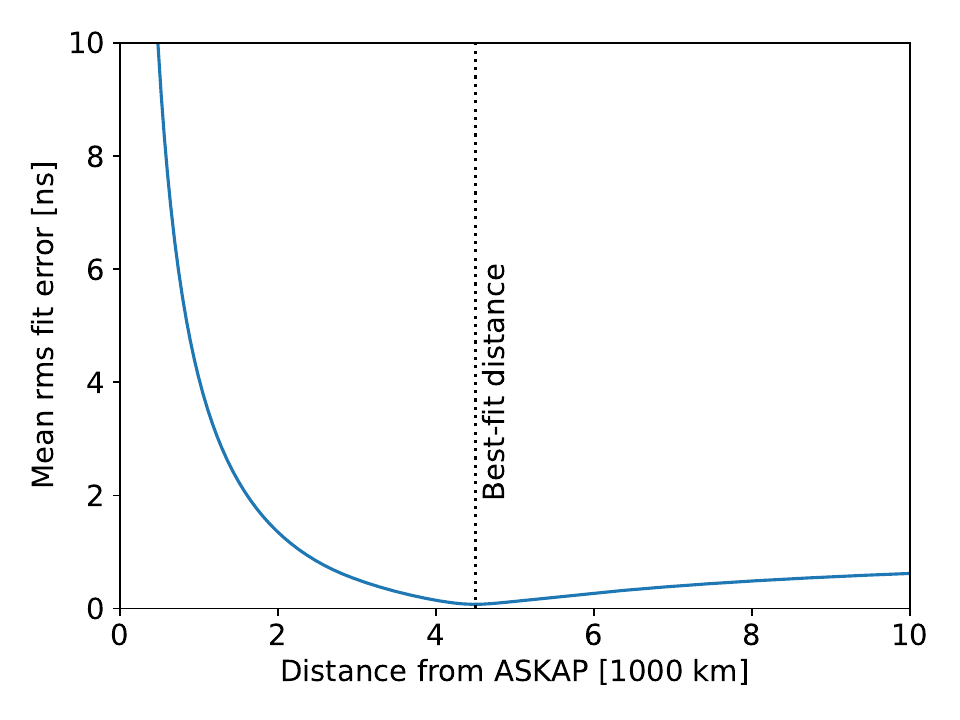}
\caption{\textbf{The best-fitting nearfield distance.} Shown is the rms of the timing residuals over all 19 antennas used in the fit. The minimum at 4500\,km corresponds to the best-fit distance for the \burst.}
\label{fig:dist_fit}
\end{figure}

To determine the event characteristics, we first measured the propagation delay between every antenna pair (baseline) independently. To do so, we made use of the far-field calibrated X- and Y- polarisation data produced by the CELEBI pipeline (that are normally used to form a "tied" beam for the purpose of studying burst structure at high time and frequency resolution. Compared to the raw voltages recorded by the ASKAP antennas, these data were gain and polarisation calibrated, and phase-offset relative to a far-field planar wavefront from the apparent source position. Additionally, artefacts due to ASKAP's over-sampled polyphase filterbank were corrected. These signals was first Fourier transformed to the time domain using a complex-to-complex inverse FFT, giving data at $1/336\,\upmu$s$ \sim 3$\,ns resolution. This revealed a raw \burst\ duration of approximately 200\,ns on \nusedant\ of the 22 antennas --- the burst was not present on three antennas, likely due to data corruption. Data from these \nusedant\ antennas were cropped about the \burst\, and correlated with antenna ak06 to extract a relative arrival time. This revealed time delays of up to 3.4\,ns on baselines over 3\,km perpendicular to the arrival direction. The arrival times were then fit with a quadratic delay, with time offset scaling as the square of the projected transverse distance about the array centre. A best-fit near-field distance of \dist\ was obtained (see Figure~\ref{fig:dist_fit}), with rms residuals of 0.075\,ns. An uncertainty of \disterr\ was determined by adding rms errors of 0.075\,ns to the time-delay data, and re-fitting for the distance.

\subsection{Event properties}
\label{sec:properties}

The identification of a near-field origin allowed CELEBI to be re-run using the delay model derived above. Hence, all burst properties reported below are derived from coherently adding all antennas to form the highest possible S/N beam. A dispersion search was applied to this data, maximising $\sum|dI/dt|^2$. No low-pass filter was applied, as is usually the case \citep{Sutinjo2023}, since structure was observed at the highest time resolution. This produced a best-fit dispersion measure of \dm, corresponding to \tecu\ (total electron content units). This agrees closely with the estimated value of slant total electron content (STEC) of $66 \pm 7$ TECU using different ionospheric models within the {\sc TECOR} package in {\sc AIPS} \citep{AIPS}, and 58.9\,TECU derived from an ionex map provided by the Centre for Orbit Determination in Europe using the ionFR package \citep{2013A&A...552A..58S}.

\begin{figure}
\centering
\includegraphics[width=\columnwidth,trim={0cm 0cm 0cm 0cm},clip]{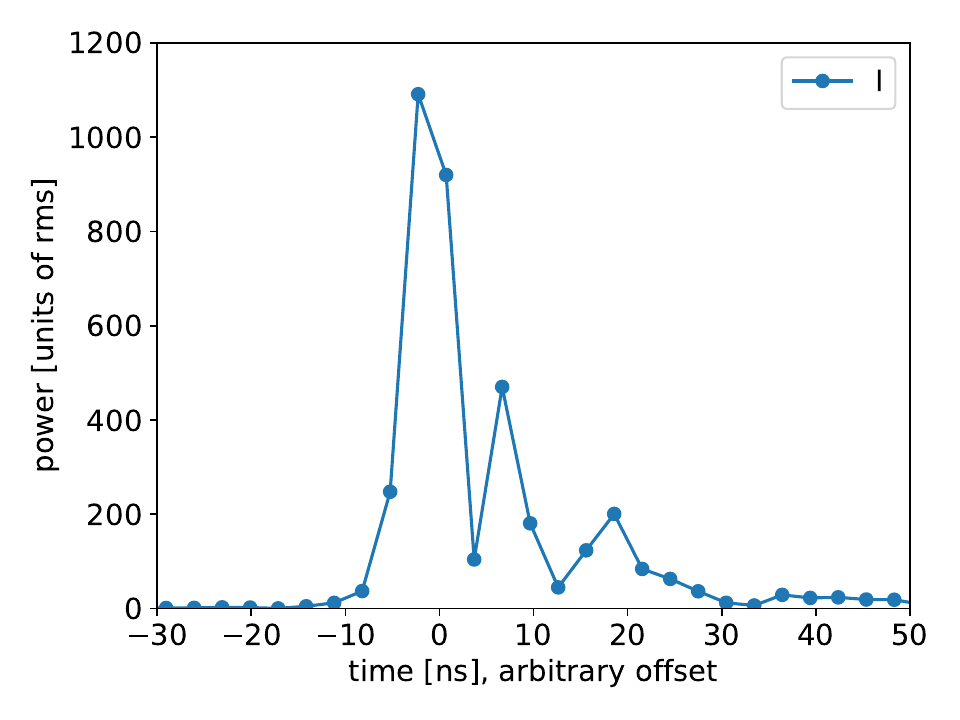}
\caption{\textbf{The dedispersed \burst.} Total burst power at Nyquist resolution (points --- lines to guide the eye only) after dedispersion and coherent addition over \nusedant\ antennas according to the nearfield fit. Note that our ability to determine the signal flux density --- and potentially also shape --- is limited by our 1-bit sampling precision (see text).}
\label{fig:nanosecond}
\end{figure}

Coherently de-dispersing the \burst\ at this DM allowed the intrinsic burst structure to be determined. The burst, plotted in Figure~\ref{fig:nanosecond}, consists of a primary impulse of \duration\ duration, with subsequent lower-intensity structure of total duration \totalduration\ (given our time resolution of 3\,ns, we cannot distinguish between a sequence of pulses of decreasing amplitude, versus a single pulse with an oscillating exponential decay). The peak power is 1090 times the off-pulse mean. While this corresponds to a peak power of \bufferflux\ given ASKAP's single-dish spectral-equivalent flux density of 1950\,Jy at our frequencies \citep{Hotan2021ASKAP}, this approximately corresponds to the limit of our dynamic range given the one-bit sampling of our buffered data. Indeed, the CRACO S/N trigger (which operates on 8-bit sampling) implies a \totalduration\ average flux density of \flux, which itself may be at the limit of the dynamic range of the ASKAP system to an impulsive event. Hence, we suspect that the primary pulse is a bandwidth-limited impulse of true duration 3\,ns or less and peak flux density of 3\,MJy or more. Assuming the \fluxmax\ value over our 336\,MHz bandwidth gives an incident power of 10$^{-11}$ W\,m$^{-2}$. At a distance of $4500$\,km, this corresponds to a source electric field strength of 280\,V\,m$^{-1}$; assuming 2\,sr of emission from a surface (see \S\ref{sec:origin}), this implies a peak power output of 400\,W (energy of 1.2\,$\upmu$\,J and assuming a 3\,ns intrinsic duration).

\begin{figure}
\centering
\includegraphics[width=\columnwidth,trim={0cm 0cm 0cm 0cm},clip]{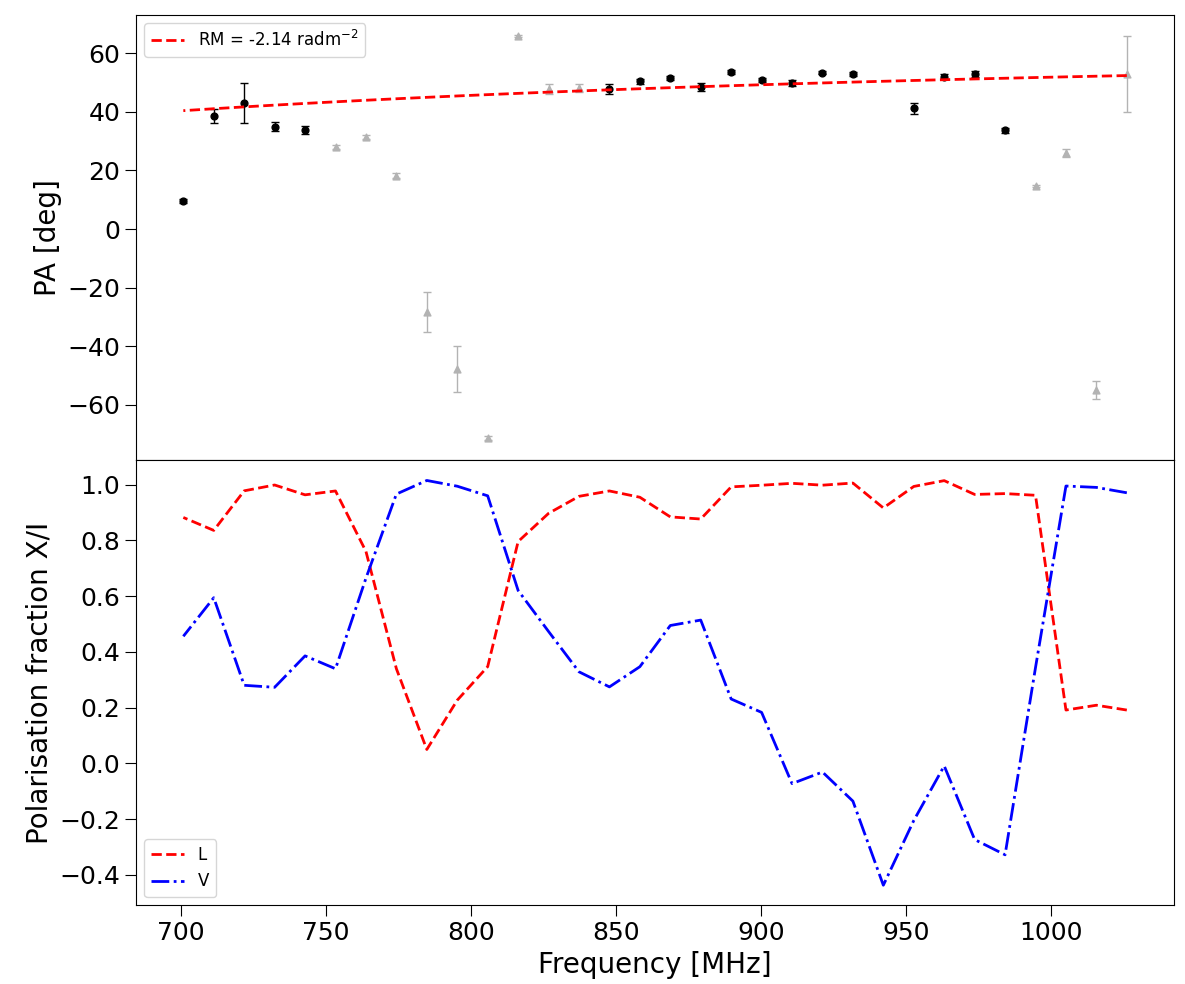}
\caption{\textbf{Polarisation properties of the \burst.} Upper panel: rotation measure (RM) fit to the position angle (PA) of the linear polarisation (black circles) as a function of frequency, with channels not used in the fit shown in grey. Lower panel: the fractional linear (L) and circular (V) polarisations as a function of frequency. Calculations were performed at 10.5\,MHz, 95.24\,ns resolution.}
\label{fig:polarisation}
\end{figure}

We determine the rotation measure (RM) of the event to be -2.14\,rad\,m$^{-2}$ --- smaller in magnitude than, but comparable to, a value of $-$3.0\,rad\,m$^{-2}$ expected from our ionospheric modelling. De-rotating the event allows us to determine its full polarisation properties, as shown in Figure~\ref{fig:polarisation}.
At most frequencies, it is purely linearly polarised, but shifts to being 100\% circularly polarised at two specific parts of the band, near 800\,MHz and 1\,GHz. Since polarisation properties are changing at inverse bandwidth resolution, our analysis forces the total polarisation fraction to be 100\% --- when polarisation properties are averaged over the burst, it is 75\% polarised.

\section{Origin of the \burst}
\label{sec:origin}

\begin{figure}
\centering
\includegraphics[width=\columnwidth,trim={0cm 0cm 0cm 0cm},clip]{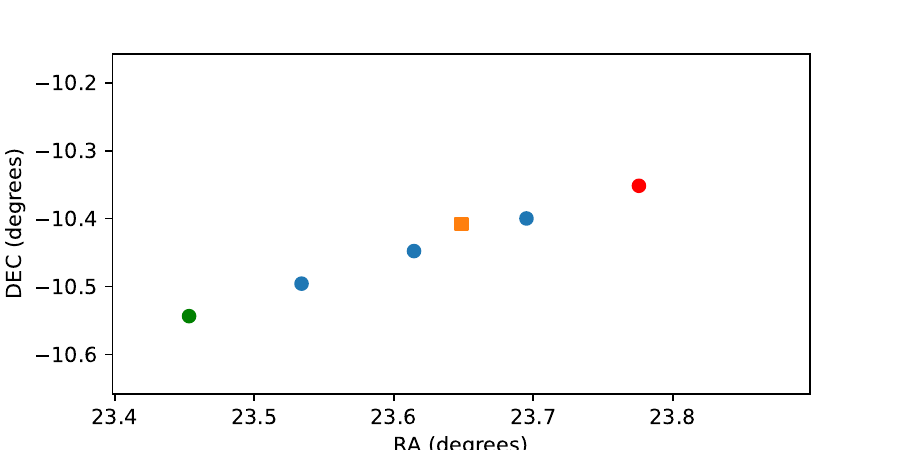}
\caption{\textbf{The match of the observed burst position and the \sat\ position} The orange square marker is the observed burst position.  The blue markers are positions of \sat, with green and red markers indicating first and last time points, respectively, separated by one second and centred on the observed burst time.}
\label{fig:match}
\end{figure}

The 4500 km distance to the burst suggests an Earth satellite as the origin.  Using the location of ASKAP, the time of the burst, and the Skyfield python module\footnote{https://rhodesmill.org/skyfield/}, we searched for a coincidence in time and position on the sky between the burst and Earth satellites.  The Two Line Element (TLE) descriptions of Earth satellite orbital parameters were downloaded from the space-track.org catalog, for all satellites in the catalog and for dates two days either side of the observation time.  Skyfield was used to propagate the orbital parameters to the observation time and calculate the Right Ascension and Declination for each satellite, as seen from the location of ASKAP.  One viable match was found, for NORAD ID 737 (\sat).\footnote{https://www.n2yo.com/satellite/?s=737} \sat\ was seen to be within 3.2\arcmin\ of the observed \burst\ position at the observation time.  Skyfield also calculates the distance between ASKAP and \sat\ at 4322 km at the observation time --- reasonably consistent with the estimated distance of \dist\ $\pm$ \disterr.  An angular distance of 3.2\arcmin\ corresponds to 4 km in the plane of the sky at a distance of 4322 km. Figure~\ref{fig:match} shows the position on the sky, as seen from ASKAP, of the observed burst location and \sat\ in a five-second window centred on the observed burst time.
We therefore conclude that this burst originated with \sat.

The \sat\ satellite was launched by NASA on 1964 Jan.\ 21, on an elliptical orbit with perigee 2091\,km, apogee 7411\,km, and inclination 46.3$^{\circ}$.\footnote{https://nssdc.gsfc.nasa.gov/nmc/spacecraft/displayTrajectory.action?id=1964-003A}
It was primarily a telecommunications satellite, with uplink frequencies between 1723.33 and 1726.67 MHz, and downlink frequencies from 4164.72\,MHz to 4174.72\,MHz \citep{RelayI}. It contained two transponders, both of which had ceased to function by 1967 June 9. It also carried three particle detection experiments --- a set of proton-electron detectors, a radiation damage experiment, and a solid-state ion chamber --- designed to study the inner Van Allen radiation belt.  A comprehensive final report on the virtually identical RELAY 1 mission can be found at https://ntrs.nasa.gov/citations/19660000937.

Operations of \sat\ ceased on September 26, 1965 \citep{NASA1965}. We therefore rule out that the \burst\ was a deliberate transmission. While re-commencement of transmissions by defunct satellites has been observed\footnote{\url{https://en.wikipedia.org/wiki/Zombie_satellite}}, a \totalduration\ pulse does not naturally correspond to any on-board system. 

We have searched 0.5\,s of data about this signal for additional sub-pulses, and find the peak power to be only $6.5$ standard deviations above the mean, consistent with random fluctuations. Since we have not yet verified the nearfield tracking of objects in our search pipeline, as few as five of the innermost antennas would coherently detect the moving satellite over the full 0.5\,s. Nonetheless, this corresponds to a limit of 2500\,Jy on associated bursts from \sat.

In \tres\ mode, CRACO has observed for 100.5\,days between 2024 May $6^{\rm th}$ and 2025 Feb.\ 19$^{\rm th}$, during which \sat\ was visible for 35 minutes. A 90\% confidence limit on bursts from \sat\ is therefore once per 15--350 minutes, and a limit of once per 44--1000 days for such bursts from all spacecraft visible to ASKAP. 

\subsection{Electrostatic discharge from a current arc}
\label{sec:ESD}

Spacecraft primarily charge through the accumulation of electrons through interactions with plasma in the space environment \citep{SpaceEnvironment}.
When sufficient voltage is achieved, electrostatic discharge (ESD) occurs, typically between nearby surfaces/materials on the spacecraft. ESD does not depend on the operational nature of the spacecraft, so ESD from a satellite decommissioned $60$\,yr ago is entirely plausible.

It has long been known that ESD causes radio frequency pulses. This has been studied in laboratory conditions \citep[][available at \url{https://ntrs.nasa.gov/citations/19860010269}]{1985jpl..rept.....L}, and through sensors mounted on spacecraft \citep[e.g.][]{PASPPlus}. Most similarly to our \burst\ detection, \citet{AreciboGPS2017} performed an observational campaign using the Arecibo ground-based radio telescope. Measurements made in the 300--350\,MHz band detected a large number of impulsive bursts from a beam centred on a GPS satellite, with the greatest having a peak flux density corresponding to 3000\,Jy. The observed arcing was attributed to charging and breakdown of the 1.5\,m solar array on the satellites, and the arcing rate (coupled to the expected damage per arcing event) was consistent with the long-term power degradation rate observed for these arrays.

The bursts seen by \citet{AreciboGPS2017} have key similarities with our event. Given the greater distance to a GPS orbit of approximately 20,000\,km, the expected flux density had the satellite been at a distance of 4300\,km would have been 65\,kJy --- significantly less than what we observe. However, as a very early spacecraft, \sat\ may have been constructed from materials with higher resistivities capable of holding greater charge and hence producing stronger ESD events \citep{MitigatingChargingEffects}. Furthermore, the pulse shape plotted in Figure~7 of their work exhibits a similar secondary peak of magnitude $\sim 25$\% of the peak power. However, the typical observed burst duration for Arecibo bursts was 40-- 50\,$\upmu$s, i.e.\ at least 1000 times longer than our \burst. A detailed analysis of arcing from photovoltaic cells in laboratory experiments with $0.5$\,ns resolution by \citet{FergusonSpectral2022} reveals arc currents varying at rates of up to $200$\,MHz, but the currents undergo many oscillations, and do not produce a single spike of emission. The typical rise-time of arcing currents from dielectric plates to conductive substrates is approximately 3\,ns, with a total duration of 10\,ns, and would thus be capable of producing the observed emission, though other materials also experience extremely short rise-times \citep{ESDtests}. While the above analyses do not examine the polarisation signature, it is expected that arcing produces polarisation in the observer--current plane \citep{ArcDischargeTheory}. A linear current would then produce linear polarisation.

We can find no specific trigger for such an ESD. \sat's orbit passes through the inner Van Allen belt, though at the time of the \burst, it was at a significantly lower altitude. While the Sun was near the peak of the solar cycle, it was unusually inactive, with a global $K_p$ index of $1$ at the time of the event \citep[][ available at \url{https://kp.gfz-potsdam.de/en/}]{2021SpWea..1902641M}, so that anomalously high charging due to geomagnetic activity can be excluded. Hence, such bursts from \sat\ may be a regular occurrence under normal conditions.

\subsection{A micrometeoroid impact scenario}
\label{sec:mm}

It is also possible that the \burst\ was triggered by a micrometeoroid impact, which can result in charging events due to the plasma cloud generated upon impact, or discharge due to this plasma increasing the conductivity of the space environment \citep{2013ITPS...41.3545G}.
Micrometeoroid impacts can also produce direct radio-frequency emission. Laboratory experiments simulating these impacts on a variety of materials (at typically less than 10\,km\,s$^{-1}$) have detected pulses lasting 2\,ns \citep{micrometeoroidobs}, with the suggested emission mechanism being due to discharge over the cracks produced in a material from the impact. These pulses were found to occur in groups however, with individual pulses separated by 17\,ns to 40\,$\upmu$s. \citet{LI2023104619} found qualitatively similar behaviour, with emission being strongest at a very low frequency of 1\,kHz, but with a significant frequency peak at 10\,MHz. However, very narrow spikes ($\ll 100$\,ns) in the time domain were also present, suggesting that higher-frequency emission, perhaps up to the ASKAP observation band, is also possible.

\citet{FletcherPhD} have performed smooth-particle hydrodynamics simulations of impacts up to 72\,km\,s$^{-1}$, finding field strengths of 10\,V/m at 40\,cm distance for a 1\,ng ($4.6\,\upmu$m) impactor travelling at 20\,km\,s$^{-1}$. The field was predominantly in a single direction, implying that linearly polarised emission would be produced. The author notes that this emission is direct emission from the plasma, which is produced for impacts above approximately 18\,km\,s$^{-1}$ when the plasma is fully ionised, and that the peak electric field strength tends to saturate at higher velocities/impactor masses. However, if the linear dimensions of the plasma cloud scale with the cube root of the impactor mass, the far-field strength will also scale proportionally, so that a $22\,\upmu$g micrometeoroid could produce the observed field strength for our \burst. Estimates of the micrometeoroid flux give a rate of $2 \cdot 10^{-9}$\,m$^{-2}$\,s$^{-1}$ above this mass \citep{micrometeoroidobs}. Assuming an average of one satellite with 1\,m$^2$ cross-section in ASKAP's field of view at any given moment (at a characteristic distance of 1000\,km, ASKAP's 30\,deg$^2$ FOV covers an area of approximately 10,000\,km$^2$, giving 0.2 satellites per FOV when compared to $\mathcal{O}\sim 10^4$ satellites spread over $4 \pi R_{E}^2 \sim 5 \cdot 10^8$\,km for an Earth radius of $R_E = 6,378$\,km), this translates to a probability of 1.7\% in the 100.5 days of CRACO observation.

We therefore consider emission from a micrometeoroid impact to be a plausible explanation for the \burst, but favour the ESD scenario due to similarities with ESD events observed with Arecibo, and because ESD may be more likely from an older satellite. We note that neither emission mechanism explains the two circularly polarised frequency bands shown in Figure~\ref{fig:polarisation}.

\section{Implications}
\label{sec:implications}

Nanosecond-scale radio impulses are commonly studied in other applications. In particular, experiments measuring radio emission from high-energy particle cascades search for impulsive signals of 1--100\,ns duration \citep{2017PrPNP..93....1S}, and regularly encounter RFI on these timescales. This makes bursts (be they due to ESD or micrometeoroid impacts) a potential impostor signal for experiments looking for radio-emission from high-energy particles.
Conversely, such astroparticle experiments could readily be made capable of detecting space-borne impulsive radio-frequency events. Cosmic ray searches with the Low Frequency Array \citep[LOFAR; ][]{LOFAR2013}, and the Auger Engineering Radio Array \citep[AERA; ][]{AERA}, operate in the 10--100\,MHz range on nanosecond timescales, and should be able to identify such events.

In addition to ASKAP, there are a large number of experiments operating in the 400\,MHz-1.5\,GHz range that search for fast radio bursts on ${\mathcal{O}}\sim$ms timescales \citep[e.g.][]{Meertrap_2023_sample,chime21cat1,dsa23cat}, many of which have relatively large fields of view that regularly observe satellites coincidentally. In particular, instruments designed for the all-sky monitoring of FRBs at GHz frequencies \citep[e.g.][]{GREX,CASPA} could, if appropriately retrofitted, be ideal for detecting such events.

\section{Conclusion}
\label{sec:conclusion}

We have observed a \totalduration\ burst of \bfmin--\bfmax\ radio emission from the non-operational \sat\ satellite. Our ability to resolve the peak of the burst, which we estimate to be $<3$\,ns duration and $>$\fluxmax, is limited by the dynamic range of the ASKAP/CRACO system. We suggest that the \burst\ originated from an electrostatic discharge (ESD) event, or potentially a micrometeoroid impact, and consider that such events may be relatively common. The \burst\ properties are consistent with those previously observed from a GPS satellite, but on a 1000 times shorter timescale. 
The observation of such a short burst at GHz frequencies is unexpected, and raises the prospect of new methods of remote sensing of arc discharges from satellites, either from retrofitting existing experiments searching for fast radio bursts or high-energy particles, or new dedicated instruments.

\section*{Acknowledgements}

This work uses data obtained from Inyarrimanha Ilgari Bundara / the CSIRO Murchison Radio-astronomy Observatory. We acknowledge the Wajarri Yamaji People as the Traditional Owners and native title holders of the Observatory site. CSIRO’s ASKAP radio telescope is part of the Australia Telescope National Facility (https://ror.org/05qajvd42). Operation of ASKAP is funded by the Australian Government with support from the National Collaborative Research Infrastructure Strategy. ASKAP uses the resources of the Pawsey Supercomputing Research Centre. Establishment of ASKAP, Inyarrimanha Ilgari Bundara, the CSIRO Murchison Radio-astronomy Observatory and the Pawsey Supercomputing Research Centre are initiatives of the Australian Government, with support from the Government of Western Australia and the Science and Industry Endowment Fund.

CRACO was funded through Australian Research Council Linkage Infrastructure Equipment, and Facilities grant LE210100107.

Part of this work was performed on the OzSTAR national facility at Swinburne University of Technology. The OzSTAR program receives funding in part from the Astronomy National Collaborative Research Infrastructure Strategy (NCRIS) allocation provided by the Australian Government, and from the Victorian Higher Education State Investment Fund (VHESIF) provided by the Victorian Government.

RMS acknowledges support through ARC Future Fellowship FT190100155 and Discovery Project DP220102305.
ATD and JJ-S acknowledges support through Australian Research Council Discovery Project DP220102305. MG is supported by the Australian Government through the Australian Research Council’s Discovery Projects funding scheme (DP210102103), and through UK STFC Grant ST/Y001117/1. MG acknowledges support from the Inter-University Institute for Data Intensive Astronomy (IDIA). IDIA is a partnership of the University of Cape Town, the University of Pretoria and the University of the Western Cape. For the purpose of open access, the author has applied a Creative Commons Attribution (CC BY) licence to any Author Accepted Manuscript version arising from this submission.

\facilities{ASKAP}

\software{{\sc numpy 2.2.1} \citep{Numpy2011}, {\sc scipy 1.15.0} \citep{SciPy2019}, {\sc astropy 7.0.0}}

\bibliography{refs}{}

\begin{thebibliography}{}
\expandafter\ifx\csname natexlab\endcsname\relax\def\natexlab#1{#1}\fi
\providecommand{\url}[1]{\href{#1}{#1}}
\providecommand{\dodoi}[1]{doi:~\href{http://doi.org/#1}{\nolinkurl{#1}}}
\providecommand{\doeprint}[1]{\href{http://ascl.net/#1}{\nolinkurl{http://ascl.net/#1}}}
\providecommand{\doarXiv}[1]{\href{https://arxiv.org/abs/#1}{\nolinkurl{https://arxiv.org/abs/#1}}}

\bibitem[{{Allen}(2010)}]{2010SpWea...8.6008A}
{Allen}, J. 2010, Space Weather, 8, S06008, \dodoi{10.1029/2010SW000588}

\bibitem[{{CHIME/FRB Collaboration} {et~al.}(2021){CHIME/FRB Collaboration}, {Amiri}, {Andersen}, {Bandura}, {Berger}, {Bhardwaj}, {Boyce}, {Boyle}, {Brar}, {Breitman}, {Cassanelli}, {Chawla}, {Chen}, {Cliche}, {Cook}, {Cubranic}, {Curtin}, {Deng}, {Dobbs}, {Dong}, {Eadie}, {Fandino}, {Fonseca}, {Gaensler}, {Giri}, {Good}, {Halpern}, {Hill}, {Hinshaw}, {Josephy}, {Kaczmarek}, {Kader}, {Kania}, {Kaspi}, {Landecker}, {Lang}, {Leung}, {Li}, {Lin}, {Masui}, {McKinven}, {Mena-Parra}, {Merryfield}, {Meyers}, {Michilli}, {Milutinovic}, {Mirhosseini}, {M{\"u}nchmeyer}, {Naidu}, {Newburgh}, {Ng}, {Patel}, {Pen}, {Petroff}, {Pinsonneault-Marotte}, {Pleunis}, {Rafiei-Ravandi}, {Rahman}, {Ransom}, {Renard}, {Sanghavi}, {Scholz}, {Shaw}, {Shin}, {Siegel}, {Sikora}, {Singh}, {Smith}, {Stairs}, {Tan}, {Tendulkar}, {Vanderlinde}, {Wang}, {Wulf}, \& {Zwaniga}}]{chime21cat1}
{CHIME/FRB Collaboration}, {Amiri}, M., {Andersen}, B.~C., {et~al.} 2021, \apjs, 257, 59, \dodoi{10.3847/1538-4365/ac33ab}

\bibitem[{{Connor} {et~al.}(2021){Connor}, {Shila}, {Kulkarni}, {Flygare}, {Hallinan}, {Li}, {Lu}, {Ravi}, \& {Weinreb}}]{GREX}
{Connor}, L., {Shila}, K.~A., {Kulkarni}, S.~R., {et~al.} 2021, \pasp, 133, 075001, \dodoi{10.1088/1538-3873/ac0bcc}

\bibitem[{{Damas} \& {Robiscoe}(1988)}]{ArcDischargeTheory}
{Damas}, M.~C., \& {Robiscoe}, R.~T. 1988, Journal of Applied Physics, 64, 566, \dodoi{10.1063/1.341971}

\bibitem[{{DeForest}(1972)}]{SpacecraftCharging1972}
{DeForest}, S.~E. 1972, \jgr, 77, 651, \dodoi{10.1029/JA077i004p00651}

\bibitem[{{Ferguson} {et~al.}(2016){Ferguson}, {Crabtree}, {White}, \& {Vayner}}]{FergusenGPS2016}
{Ferguson}, D., {Crabtree}, P., {White}, S., \& {Vayner}, B. 2016, Journal of Spacecraft and Rockets, 53, 464, \dodoi{https://doi.org/10.2514/1.A33438}

\bibitem[{{Ferguson} {et~al.}(2017){Ferguson}, {White}, {Rast}, {Balasubramaniam}, {Thompson}, {Suszcynsky}, \& {Holeman}}]{AreciboGPS2017}
{Ferguson}, D., {White}, S., {Rast}, R., {et~al.} 2017, Journal of Spacecraft and Rockets, 54, 566, \dodoi{10.2514/1.A33724}

\bibitem[{{Ferguson} {et~al.}(2014){Ferguson}, {Murray-Krezan}, {Barton}, {Dennison}, \& {Gregory}}]{2014JSpRo..51.1907F}
{Ferguson}, D.~C., {Murray-Krezan}, J., {Barton}, D.~A., {Dennison}, J.~R., \& {Gregory}, S.~A. 2014, Journal of Spacecraft and Rockets, 51, 1907, \dodoi{10.2514/1.A32958}

\bibitem[{{Ferguson} {et~al.}(2022){Ferguson}, {Perillat}, \& {Vayner}}]{FergusonSpectral2022}
{Ferguson}, D.~C., {Perillat}, P., \& {Vayner}, B. 2022, Journal of the Astronautical Sciences, 69, 139, \dodoi{10.1007/s40295-021-00295-8}

\bibitem[{{Fletcher}(2015)}]{FletcherPhD}
{Fletcher}, A. 2015, PhD thesis, Stanford University, Department of Aeronautics and Astronautics

\bibitem[{{Garrett} \& { Whittlesey}(2012)}]{MitigatingChargingEffects}
{Garrett}, H.~B., \& { Whittlesey}, A.~C. 2012, Guide to Mitigating Spacecraft Charging Effects (John Wiley \& Sons, Ltd), \dodoi{https://doi.org/10.1002/9781118241400.ch1}

\bibitem[{{Garrett} \& {Close}(2013)}]{2013ITPS...41.3545G}
{Garrett}, H.~B., \& {Close}, S. 2013, IEEE Transactions on Plasma Science, 41, 3545, \dodoi{10.1109/TPS.2013.2286181}

\bibitem[{{Goddard Space Flight Centre}(1965)}]{RelayI}
{Goddard Space Flight Centre}. 1965, {Final Report on the Relay I Program. NASA SP-76}, Vol.~76 (National Aeronautics and Space Administration).
\newblock \url{https://ntrs.nasa.gov/api/citations/19660000937/downloads/19660000937.pdf}

\bibitem[{{Greisen}(2003)}]{AIPS}
{Greisen}, E.~W. 2003, in Astrophysics and Space Science Library, Vol. 285, Information Handling in Astronomy - Historical Vistas, ed. A.~{Heck}, 109, \dodoi{10.1007/0-306-48080-8_7}

\bibitem[{{Guidice} {et~al.}(1992){Guidice}, {Davis}, {Curtis}, {Ferguson}, {Hastings}, {Knight}, {Marvin}, {Ray}, {Severance}, {Soldi}, \& {Van Riet}}]{PASPPlus}
{Guidice}, D.~A., {Davis}, V.~A., {Curtis}, H.~B., {et~al.} 1992, in NASA Conference Publication, Vol. 3127, Fifth Annual Workshop on Space Operations Applications and Research (SOAR 1991), ed. K.~{Krishen}, 662

\bibitem[{{Hotan} {et~al.}(2021){Hotan}, {Bunton}, {Chippendale}, {Whiting}, {Tuthill}, {Moss}, {McConnell}, {Amy}, {Huynh}, {Allison}, {Anderson}, {Bannister}, {Bastholm}, {Beresford}, {Bock}, {Bolton}, {Chapman}, {Chow}, {Collier}, {Cooray}, {Cornwell}, {Diamond}, {Edwards}, {Feain}, {Franzen}, {George}, {Gupta}, {Hampson}, {Harvey-Smith}, {Hayman}, {Heywood}, {Jacka}, {Jackson}, {Jackson}, {Jeganathan}, {Johnston}, {Kesteven}, {Kleiner}, {Koribalski}, {Lee-Waddell}, {Lenc}, {Lensson}, {Mackay}, {Mahony}, {McClure-Griffiths}, {McConigley}, {Mirtschin}, {Ng}, {Norris}, {Pearce}, {Phillips}, {Pilawa}, {Raja}, {Reynolds}, {Roberts}, {Roxby}, {Sadler}, {Shields}, {Schinckel}, {Serra}, {Shaw}, {Sweetnam}, {Troup}, {Tzioumis}, {Voronkov}, \& {Westmeier}}]{Hotan2021ASKAP}
{Hotan}, A.~W., {Bunton}, J.~D., {Chippendale}, A.~P., {et~al.} 2021, \pasa, 38, e009, \dodoi{10.1017/pasa.2021.1}

\bibitem[{{James} {et~al.}(1994){James}, Norton, \& Alexander}]{SpaceEnvironment}
{James}, B.~F., Norton, O., \& Alexander, M.~B. 1994, {The Natural Space Environment: Effects on Spacecraft}, Vol. NASA Reference Publication 1350 (National Aeronautics and Space Administration).
\newblock \url{https://ntrs.nasa.gov/api/citations/19950019455/downloads/19950019455.pdf}

\bibitem[{{Jankowski} {et~al.}(2023){Jankowski}, {Bezuidenhout}, {Caleb}, {Driessen}, {Malenta}, {Morello}, {Rajwade}, {Sanidas}, {Stappers}, {Surnis}, {Barr}, {Chen}, {Kramer}, {Wu}, {Buchner}, {Serylak}, \& {Prochaska}}]{Meertrap_2023_sample}
{Jankowski}, F., {Bezuidenhout}, M.~C., {Caleb}, M., {et~al.} 2023, arXiv e-prints, arXiv:2302.10107, \dodoi{10.48550/arXiv.2302.10107}

\bibitem[{{Jet Propulsion Laboratory}(1999)}]{ESDtests}
{Jet Propulsion Laboratory}. 1999, Electrostatics discharge (ESD) test practices, Tech. Rep. PRACTICE NO. PT-TE-1414, National Aeronautics and Space Administration.
\newblock \url{https://llis.nasa.gov/lesson/777}

\bibitem[{{Law} {et~al.}(2023){Law}, {Sharma}, {Ravi}, {Chen}, {Catha}, {Connor}, {Faber}, {Hallinan}, {Harnach}, {Hellbourg}, {Hobbs}, {Hodge}, {Hodges}, {Lamb}, {Rasmussen}, {Sherman}, {Shi}, {Simard}, {Squillace}, {Weinreb}, {Woody}, \& {Yadlapalli}}]{dsa23cat}
{Law}, C.~J., {Sharma}, K., {Ravi}, V., {et~al.} 2023, arXiv e-prints, arXiv:2307.03344, \dodoi{10.48550/arXiv.2307.03344}

\bibitem[{{Leach} \& {Alexander}(1995)}]{1995STIN...9611547L}
{Leach}, R.~D., \& {Alexander}, M.~B. 1995, {Failures and anomalies attributed to spacecraft charging}

\bibitem[{{Leung}(1985)}]{1985jpl..rept.....L}
{Leung}, P. 1985, {Characterization of EMI generated by the discharge of a VOLT solar array}, Final Report Jet Propulsion Lab., California Inst. of Tech., Pasadena.

\bibitem[{Li {et~al.}(2023)Li, Zhang, Han, Cai, \& Tao}]{LI2023104619}
Li, H., Zhang, L., Han, J., Cai, M., \& Tao, M. 2023, International Journal of Impact Engineering, 178, 104619, \dodoi{https://doi.org/10.1016/j.ijimpeng.2023.104619}

\bibitem[{{Luo} {et~al.}(2024){Luo}, {Ekers}, {Hobbs}, {Dunning}, {James}, {Lower}, {Gupta}, {Zic}, {Sokolowski}, {Phillips}, {Deller}, \& {Staveley-Smith}}]{CASPA}
{Luo}, R., {Ekers}, R., {Hobbs}, G., {et~al.} 2024, \pasa, 41, e109, \dodoi{10.1017/pasa.2024.108}

\bibitem[{Maki {et~al.}(2005)Maki, Soma, Takano, Fujiwara, \& Yamori}]{micrometeoroidobs}
Maki, K., Soma, E., Takano, T., Fujiwara, A., \& Yamori, A. 2005, Journal of Applied Physics, 97, 104911, \dodoi{10.1063/1.1896092}

\bibitem[{{Matzka} {et~al.}(2021){Matzka}, {Stolle}, {Yamazaki}, {Bronkalla}, \& {Morschhauser}}]{2021SpWea..1902641M}
{Matzka}, J., {Stolle}, C., {Yamazaki}, Y., {Bronkalla}, O., \& {Morschhauser}, A. 2021, Space Weather, 19, e2020SW002641, \dodoi{10.1029/2020SW002641}

\bibitem[{{NASA Historical Staff, Office of Policy Analysis}(1966)}]{NASA1965}
{NASA Historical Staff, Office of Policy Analysis}. 1966, {Astronautics and Aeronautics, 1965: Chronology on Science Technology, and Policy. NASA SP-4006}, Vol. 4006 (National Aeronautics and Space Administration).
\newblock \url{https://www.nasa.gov/wp-content/uploads/2023/04/1965.pdf?emrc=ee9c46}

\bibitem[{{Schellart} {et~al.}(2013){Schellart}, {Nelles}, {Buitink}, {Corstanje}, {Enriquez}, {Falcke}, {Frieswijk}, {H{\"o}randel}, {Horneffer}, {James}, {Krause}, {Mevius}, {Scholten}, {ter Veen}, {Thoudam}, {van den Akker}, {Alexov}, {Anderson}, {Avruch}, {B{\"a}hren}, {Beck}, {Bell}, {Bennema}, {Bentum}, {Bernardi}, {Best}, {Bregman}, {Breitling}, {Brentjens}, {Broderick}, {Br{\"u}ggen}, {Ciardi}, {Coolen}, {de Gasperin}, {de Geus}, {de Jong}, {de Vos}, {Duscha}, {Eisl{\"o}ffel}, {Fallows}, {Ferrari}, {Garrett}, {Grie{\ss}meier}, {Grit}, {Hamaker}, {Hassall}, {Heald}, {Hessels}, {Hoeft}, {Holties}, {Iacobelli}, {Juette}, {Karastergiou}, {Klijn}, {Kohler}, {Kondratiev}, {Kramer}, {Kuniyoshi}, {Kuper}, {Maat}, {Macario}, {Mann}, {Markoff}, {McKay-Bukowski}, {McKean}, {Miller-Jones}, {Mol}, {Mulcahy}, {Munk}, {Nijboer}, {Norden}, {Orru}, {Overeem}, {Paas}, {Pandey-Pommier}, {Pizzo}, {Polatidis}, {Renting}, {Romein}, {R{\"o}ttgering}, {Schoenmakers}, {Schwarz}, {Sluman}, {Smirnov}, {Sobey}, {Stappers},
  {Steinmetz}, {Swinbank}, {Tang}, {Tasse}, {Toribio}, {van Leeuwen}, {van Nieuwpoort}, {van Weeren}, {Vermaas}, {Vermeulen}, {Vocks}, {Vogt}, {Wijers}, {Wijnholds}, {Wise}, {Wucknitz}, {Yatawatta}, {Zarka}, \& {Zensus}}]{LOFAR2013}
{Schellart}, P., {Nelles}, A., {Buitink}, S., {et~al.} 2013, \aap, 560, A98, \dodoi{10.1051/0004-6361/201322683}

\bibitem[{{Schr{\"o}der}(2017)}]{2017PrPNP..93....1S}
{Schr{\"o}der}, F.~G. 2017, Progress in Particle and Nuclear Physics, 93, 1, \dodoi{10.1016/j.ppnp.2016.12.002}

\bibitem[{{Scott} {et~al.}(2023){Scott}, {Cho}, {Day}, {Deller}, {Glowacki}, {Gourdji}, {Bannister}, {Bera}, {Bhandari}, {James}, \& {Shannon}}]{Scott2023_CELEBI}
{Scott}, D.~R., {Cho}, H., {Day}, C.~K., {et~al.} 2023, Astronomy and Computing, 44, 100724, \dodoi{10.1016/j.ascom.2023.100724}

\bibitem[{{Shannon} {et~al.}(2024){Shannon}, {Bannister}, {Bera}, {Bhandari}, {Day}, {Deller}, {Dial}, {Dobie}, {Ekers}, {Fong}, {Glowacki}, {Gordon}, {Gourdji}, {Jaini}, {James}, {Kumar}, {Mahony}, {Marnoch}, {Muller}, {Prochaska}, {Qiu}, {Ryder}, {Sadler}, {Scott}, {Tejos}, {Uttarkar}, \& {Wang}}]{shannon24craftics}
{Shannon}, R.~M., {Bannister}, K.~W., {Bera}, A., {et~al.} 2024, arXiv e-prints, arXiv:2408.02083, \dodoi{10.48550/arXiv.2408.02083}

\bibitem[{{Sotomayor-Beltran} {et~al.}(2013){Sotomayor-Beltran}, {Sobey}, {Hessels}, {de Bruyn}, {Noutsos}, {Alexov}, {Anderson}, {Asgekar}, {Avruch}, {Beck}, {Bell}, {Bell}, {Bentum}, {Bernardi}, {Best}, {Birzan}, {Bonafede}, {Breitling}, {Broderick}, {Brouw}, {Br{\"u}ggen}, {Ciardi}, {de Gasperin}, {Dettmar}, {van Duin}, {Duscha}, {Eisl{\"o}ffel}, {Falcke}, {Fallows}, {Fender}, {Ferrari}, {Frieswijk}, {Garrett}, {Grie{\ss}meier}, {Grit}, {Gunst}, {Hassall}, {Heald}, {Hoeft}, {Horneffer}, {Iacobelli}, {Juette}, {Karastergiou}, {Keane}, {Kohler}, {Kramer}, {Kondratiev}, {Koopmans}, {Kuniyoshi}, {Kuper}, {van Leeuwen}, {Maat}, {Macario}, {Markoff}, {McKean}, {Mulcahy}, {Munk}, {Orru}, {Paas}, {Pandey-Pommier}, {Pilia}, {Pizzo}, {Polatidis}, {Reich}, {R{\"o}ttgering}, {Serylak}, {Sluman}, {Stappers}, {Tagger}, {Tang}, {Tasse}, {ter Veen}, {Vermeulen}, {van Weeren}, {Wijers}, {Wijnholds}, {Wise}, {Wucknitz}, {Yatawatta}, \& {Zarka}}]{2013A&A...552A..58S}
{Sotomayor-Beltran}, C., {Sobey}, C., {Hessels}, J.~W.~T., {et~al.} 2013, \aap, 552, A58, \dodoi{10.1051/0004-6361/201220728}

\bibitem[{{Sutinjo} {et~al.}(2023){Sutinjo}, {Scott}, {James}, {Glowacki}, {Bannister}, {Cho}, {Day}, {Deller}, {Perrett}, \& {Shannon}}]{Sutinjo2023}
{Sutinjo}, A.~T., {Scott}, D.~R., {James}, C.~W., {et~al.} 2023, \apj, 954, 37, \dodoi{10.3847/1538-4357/ace774}

\bibitem[{{The Pierre Auger Collaboration}(2016)}]{AERA}
{The Pierre Auger Collaboration}. 2016, Journal of Instrumentation, 11, P01018, \dodoi{10.1088/1748-0221/11/01/P01018}

\bibitem[{{van der Walt} {et~al.}(2011){van der Walt}, {Colbert}, \& {Varoquaux}}]{Numpy2011}
{van der Walt}, S., {Colbert}, S.~C., \& {Varoquaux}, G. 2011, Computing in Science and Engineering, 13, 22, \dodoi{10.1109/MCSE.2011.37}

\bibitem[{{Virtanen} {et~al.}(2020)}]{SciPy2019}
{Virtanen}, P., {et~al.} 2020, Nature Methods, 17, 261, \dodoi{10.1038/s41592-019-0686-2}

\bibitem[{{Wang} {et~al.}(2024){Wang}, {Bannister}, {Gupta}, {Deng}, {Pilawa}, {Tuthill}, {Bunton}, {Flynn}, {Glowacki}, {Jaini}, {Lee}, {Lenc}, {Lucero}, {Paek}, {Radhakrishnan}, {Thyagarajan}, {Uttarkar}, {Wang}, {Bhat}, {James}, {Moss}, {Murphy}, {Reynolds}, {Shannon}, {Spitler}, {Tzioumis}, {Caleb}, {Deller}, {Gordon}, {Marnoch}, {Ryder}, {Simha}, {Anderson}, {Ball}, {Brodrick}, {Cooray}, {Gupta}, {Hayman}, {Ng}, {Pearce}, {Phillips}, {Voronkov}, \& {Westmeier}}]{Andy2024CRACO}
{Wang}, Z., {Bannister}, K.~W., {Gupta}, V., {et~al.} 2024, arXiv e-prints, arXiv:2409.10316, \dodoi{10.48550/arXiv.2409.10316}

\end{thebibliography}
\bibliographystyle{aasjournal}

\end{document}